\documentclass[a4paper,fleqn,usenatbib]{mnras}

\usepackage{newtxtext,newtxmath}
\usepackage[T1]{fontenc}
\usepackage{ae,aecompl}
\usepackage{graphicx}	% Including figure files
\usepackage{amsmath}	% Advanced maths commands
\usepackage{amssymb}	% Extra maths symbols

\def\widA{WASP-180A}
\def\widB{WASP-180B}
\def\widb{WASP-180Ab}
\def\prot{$P_{\rm rot}$}
\def\porb{$P_{\rm orb}$}
\def\teff{$T_{\rm eff}$}
\def\logg{$\log{g_{*}}$}
\def\vsini{$v \sin i_{\star}$}

\title[WASP-180Ab]{WASP-180Ab: Doppler tomography of an hot Jupiter orbiting the primary star in a visual binary}

\author[L. Temple et al.]{
L.Y. Temple,$^{1}$\thanks{E-mail: l.y.temple@keele.ac.uk}
C. Hellier$^{1}$,
D.R. Anderson$^{1}$,
K. Barkaoui$^{7,2}$,
F. Bouchy$^{3}$,\newauthor
D.J.A. Brown$^{4,5}$,
A. Burdanov$^{7}$,
A. Collier Cameron$^{6}$,
L. Delrez$^{8}$,
E. Ducrot$^{7}$,\newauthor
D. Evans$^{1}$,
M. Gillon$^{7}$,
E. Jehin$^{7}$,
M. Lendl$^{3}$,
P.F.L. Maxted$^{1}$,
J. McCormac$^{4}$,\newauthor
C. Murray$^{8}$,
L. D. Nielsen$^{3}$,
F. Pepe$^{3}$,
D. Pollacco$^{4,5}$,
D. Queloz$^{8}$,
D. S\'egransan$^{3}$,\newauthor
B. Smalley$^{1}$,
S. Thompson$^{8}$,
A.H.M.J. Triaud$^{9}$,
O.D. Turner$^{3,1}$,
S. Udry$^{3}$,
R.G. West$^{4}$,\newauthor
B. Zouhair$^{2}$
\\
$^{1}$Astrophysics Group, Keele University, Staffordshire, ST5 5BG, UK\\
$^{2}$Ouka\"{\i}mden Observatory, High Energy Physics and Astrophysics Laboratory, Cadi Ayyad University, Marrakech, Morocco\\
$^{3}$Observatoire astronomique de l'Universit\'e de Gen\`eve 51 ch. des Maillettes, 1290 Sauverny, Switzerland\\
$^{4}$Department of Physics, University of Warwick, Gibbet Hill Road, Coventry, CV4 7AL, UK\\
$^{5}$Centre for Exoplanets and Habitability, University of Warwick, Gibbet Hill Road, Coventry CV4 7AL, UK\\
$^{6}$SUPA, School of Physics and Astronomy, University of St.\ Andrews, North Haugh,  Fife, KY16 9SS, UK\\
$^{7}$Space sciences, Technologies and Astrophysics Research (STAR) Institute, Universit{\'e} de Li{\`e}ge, All{\'e}e du 6 Ao{\^u}t 17, 4000 Li{\`e}ge, Belgium\\
$^{8}$Cavendish Laboratory, J J Thomson Avenue, Cambridge, CB3 0HE, UK\\
$^{9}$School of Physics \& Astronomy, University of Birmingham, Edgbaston, Birmingham, B15 2TT, UK\\
}
\date{Accepted XXX. Received YYY; in original form ZZZ}

\pubyear{2019}

\begin{document}
\label{firstpage}
\pagerange{\pageref{firstpage}--\pageref{lastpage}}
\maketitle

\begin{abstract}
We report the discovery and characterisation of \widb, a hot Jupiter confirmed by the detection of its Doppler shadow and by measuring its mass using radial velocities. We find the 0.9\,$\pm$\,0.1\,$M_{\rm Jup}$, 1.24\,$\pm$\,0.04\,$R_{\rm Jup}$ planet to be in a misaligned, retrograde orbit around an F7 star with \teff\,=\,6500\,K and a moderate rotation speed of \vsini\,=\,19.9\,km\,s$^{-1}$. The host star is the primary of a $V$\,=\,10.7 binary, where a secondary separated by $\sim$5$''$ ($\sim$1200\,AU) contributes $\sim$\,30\% of the light. \widb\ therefore adds to a small sample of transiting hot Jupiters known in binary systems.  A 4.6-day modulation seen in the WASP data is likely to be the rotational modulation of the companion star, \widB.
\end{abstract}


\section{Introduction}
In the age of high resolution spectrographs, the possibilities for detailed characterisation of exoplanets are expanding. We are able to map the motion of a hot Jupiter across the disc of its host star as it transits, via a method called Doppler tomography. This method consists of the direct detection of distortions to stellar line profiles that occur due to the occultation of a portion of the stellar disc by a smaller orbiting body, a phenomenon called the Rossiter-McLaughlin (RM) effect \citep[e.g.][]{2010MNRAS.403..151C,2018AJ....155...35S,2019AJ....157..141T}. In mapping the motion of this distortion as a function of phase, it is possible to determine the current projected spin-orbit misalignment angle, $\lambda$, measured as the apparent angle between the stellar rotation axis and the normal to the orbital plane of the planet. Knowledge of $\lambda$ gives insight to the dynamical history of the system. Doppler tomography is especially suited to hotter targets which elude radial velocity characterisation due to a lack of spectral lines. It can also reveal the effects of internal stellar motions on the surface of a star, such as differential rotation and convection \citep{2016ApJ...819...67C} and stellar pulsations \citep[see, e.g. ][]{2017MNRAS.471.2743T}.

In this work we present a newly discovered hot Jupiter, \widb, transiting the primary star of a visual binary in a misaligned, retrograde orbit.

This discovery may be in line with theories surrounding Lidov-Kozai oscillations being responsible for the high obliquities seen in some hot-Jupiter systems \cite[e.g.][]{2016MNRAS.456.3671A,2017MNRAS.465.3927S}. It has long been theorized that a distant stellar companion can induce such oscillations in a Jupiter's orbit, leading to high-eccentricity migration of the planet which produces a misaligned, short-period orbit. This would then be followed by realignment of the host star with the planet's orbit via tidal dissipation, an effect that would be less efficient for stellar hosts lacking convective envelopes, and thus this theory is consistent with the observed tendency of systems with stars hotter than $\sim$\,6250\,K being more likely to have planetary orbits which are misaligned with respect to the stellar rotation axis \citep{2010ApJ...718L.145W,2012ApJ...757...18A}.

Through the works of \citet{2015ApJ...800..138N,2015ApJ...814..148P,2016ApJ...827....8N,2018A&A...610A..20E}, for example, we now know that a large portion of the known planet population consists of systems containing lower mass stellar companions, with \citet{2016ApJ...827....8N} concluding that 47\%\,$\pm$\,7\% of hot Jupiters have stellar companions at separations of 50--2000\,AU. Meanwhile, \citet{2018A&A...610A..20E} show that there is a dearth of planets in wide binary systems with stars of similar mass. It should be noted, however, that this finding is at least partially a selection bias: in systems with stars of similar mass and thus brightness, as in the case of WASP-180, the light from the planet hosting star is significantly diluted in the light of the other star, reducing the apparent transit depth and making detection via the transit method more difficult. Now, \widb\ adds to a small group of known hot Jupiters in near-equal mass stellar binaries.

\section{Data and Observations}
\label{sec:obs}
WASP-180 is a known binary, listed as WDS 08136-0159 in the Washington Double Star Catalogue \citep{2001AJ....122.3466M}, with the two stars having {\it Gaia}  magnitudes of 10.9 and 11.8. {\it Gaia} DR2 confirms the two stars to have the same parallax and proper motions, and we calculate the angular separation to be 4.854$''$ \citep{2016A&A...595A...1G,2018A&A...616A...1G}. This separation is sufficient to avoid contamination in high-resolution spectroscopic observations of the system.

We observed \widA\ from November 2009 to March 2012 using the SuperWASP-North telescope \citep{2006PASP..118.1407P} located at the Roque de los Muchachos Observatory in La Palma, as well as the WASP-South telescope \citep{2011EPJWC..1101004H} located at the South African Astronomical Observatory (SAAO). The data contains light from both \widA\ and \widB.

Upon detecting a 3.4-d transit-like signal in the WASP data we obtained focused photometry with TRAPPIST-South \citep{2011Msngr.145....2J}, resolving the two stars. These data were sufficient to show that the transit is of the brighter of the two stars, WASP-180A, but were otherwise of low quality and so we exclude the lightcurve from further analysis.

We proceeded to obtain radial velocity (RV) measurements with the Euler/CORALIE \citep{2001Msngr.105....1Q} spectrograph. WASP-180A is a fast rotating F star with broad lines giving large RV errors, so the CORALIE RVs ruled out a stellar-mass transit mimic, but were not sufficient to give a measurement of the planet's mass. Thus we attempted Doppler tomography of a transit on the night of 2018 January 5 using the ESO 3.6-m/HARPS spectrograph \citep{2002Msngr.110....9P}.  Due to an auto-guiding issue three of the spectra obtained were of low signal-to-noise and were therefore discarded.  Simultaneously during this transit we observed the lightcurve using TRAPPIST-South, using an aperture including both stars. 

After tomographic confirmation of the planet we observed further follow-up lightcurves also using apertures including both stars. These were taken with TRAPPIST-North \citep{2017JPhCS.869a2073B,KBarkaoui2019AJ} at the Ouka\"{\i}mden Observatory in Morocco and the SPECULOOS-Callisto telescope \citep{2017haex.bookE.130B} at ESO Paranal Observatory. We also obtained 6 more RVs with HARPS to constrain the planet's mass.  Details of the observations used in this work are provided in Table ~\ref{table:observations}. 

The RV measurements corresponding to each of the spectra obtained are listed in Table ~\ref{table:RVs} with the corresponding bisector span (BS) measurements. These were measured from cross-correlation functions (CCFs) computed by cross-correlating the spectra using a mask matching a G2 spectral type, over a wide correlation window covering --320\,km\,s$^{-1}$ to 380\,km\,s$^{-1}$.

\begin{table}
\caption{Details of the photometric and spectroscopic observations of \widb\ carried out for this work.}
\centering
\begin{tabular}{lcc}
\hline
Telescope/Instrument & Date & Notes \\[0.5ex]
\hline
WASP-North & 2009--2011 & 8329 points \\
WASP-South & 2011--2012 & 4359 points \\
TRAPPIST-South & 2018 Jan 5 & z'. 10s exp. \\
TRAPPIST-North & 2018 Jan 12 & z'. 11s exp. \\
SPECULOOS-Callisto & 2018 Jan 22 & z'. 8s exp. \\
ESO 3.6-m/HARPS & 2018 Jan 5 & 21 spectra \\ 
 & & through transit \\
Euler/CORALIE	& 2015--2018 & 9 RVs \\
ESO 3.6-m/HARPS & 2018 Mar & 6 RVs \\ [1ex]
\hline
\end{tabular}
\label{table:observations}
\end{table}

\begin{table}
\caption{RV measurements for \widA\ taken using the CORALIE and HARPS spectrographs for this work. The values in italics are of low signal-to-noise due to an auto-guiding issue during observation.
}
\centering
\begin{tabular}{lccrc}
\hline
BJD$_{\rm TDB}$ & RV & $\sigma$$_{\rm RV}$ & BS & $\sigma$$_{\rm BS}$ \\
--2,450,000  & (km s$^{-1}$) & (km s$^{-1}$) & (km s$^{-1}$) & (km s$^{-1}$) \\ [0.5mm]
\hline
\multicolumn{5}{l}{CORALIE:} \\
7092.644031 & 28.96 & 0.05 & --0.18 & 0.10 \\
7697.849714 & 28.96 & 0.06 & 0.01 & 0.12 \\
7751.760876 & 29.02 & 0.06 & --0.29 & 0.12 \\
8077.824503 & 28.76 & 0.04 & --0.12 & 0.08 \\
8079.836204 & 28.90 & 0.05 & --0.21 & 0.10 \\
8094.796351 & 28.65 & 0.04 & --0.00 & 0.08 \\
8140.848419 & 28.90 & 0.06 & --0.15 & 0.12 \\
8212.592949 & 29.06 & 0.06 & --0.11 & 0.12 \\
8222.600891 & 28.95 & 0.07 & --0.16 & 0.14 \\
\multicolumn{5}{l}{HARPS:} \\
8198.604103 & 29.02 & 0.02 & --0.25 & 0.04 \\
8199.643111 & 28.85 & 0.02 &  --0.04 & 0.04 \\
8201.610172 & 29.01 & 0.02 & --0.07 & 0.04 \\
8202.589668 & 29.01 & 0.02 & --0.06 & 0.04 \\
8203.572959 & 28.87 & 0.02 & --0.19 & 0.04 \\
8204.571804 & 28.91 & 0.02 & --0.01 & 0.04 \\
\multicolumn{5}{l}{HARPS (2018 Jan 05):} \\
8124.596974 & 28.99 & 0.02 & --0.23 & 0.04 \\
8124.607854 & 29.05 & 0.02 & --0.21 & 0.04 \\
8124.618421 & 28.98 & 0.02 & --0.17 & 0.04 \\
8124.629200 & 28.96 & 0.02 & --0.15 & 0.04 \\
8124.640806 & 28.90 & 0.02 & --0.11 & 0.04 \\
8124.650841 & 28.76 & 0.02 & 0.03 & 0.04 \\
8124.661525 & 28.76 & 0.02 & 0.22 & 0.04 \\
8124.672312 & 28.81 & 0.02 & 0.17 & 0.04 \\
8124.683285 & 28.91 & 0.02 & --0.12 & 0.04 \\
8124.693748 & 29.02 & 0.02 & --0.31& 0.04 \\
8124.704420 & 29.10 & 0.02 & --0.48 & 0.04 \\
8124.715299 & 29.20 & 0.03 & --0.75 & 0.06 \\
8124.725971 & 29.16 & 0.03 & --0.63 & 0.06 \\
{\it 8124.736434} & {\it 29.02} & {\it 0.04} & {\it 0.10} & {\it 0.08} \\
{\it 8124.746794} & {\it 29.16} & {\it 0.04} & {\it --0.03} & {\it 0.08} \\
{\it 8124.758090} & {\it 28.35} & {\it 0.05} & {\it --1.30} & {\it 0.10} \\
8124.770313 & 28.91 & 0.02 & --0.18 & 0.04 \\
8124.780059 & 28.95 & 0.02 & --0.22 & 0.04 \\
8124.790510 & 28.91 & 0.02 & --0.19 & 0.04 \\
8124.801182 & 28.88 & 0.02 & --0.22 & 0.04 \\
8124.812062 \medskip & 28.91 & 0.03 & --0.18 & 0.06 \\
\hline
\end{tabular}
\label{table:RVs}
\end{table}

\section{Spectral analysis}
\label{sec:specanalysis}
We analysed a median-stacked HARPS spectrum created from the 18 HARPS spectra taken on the night of 2018 Jan 5, to obtain stellar parameters. We follow the methods of \citet{2013MNRAS.428.3164D} to measure \teff\,=\,6500\,$\pm$\,150\,K and \logg\,=\,4.5\,$\pm$\,0.2\,dex. We measure \vsini\,=\,18.3\,$\pm$\,1.1\,km\,s$^{-1}$ by assuming a microturbulence value of $v_{\rm mic}$\,=\,1.5\,km\,s$^{-1}$ from the calibration of \citet{2010MNRAS.405.1907B} and a macroturbulence value of 5.8\,km\,s$^{-1}$ extrapolated from the calibrations of \citet{2014MNRAS.444.3592D}, which is valid for stars up to 6400\,K. We also measure the metallicity as [Fe/H]\,=\,0.09$\pm$0.19, and finally, use the MKCLASS program \citep{2014AJ....147...80G} to obtain a spectral type of F7 V.

\section{The distant co-moving companion}
\label{sec:thebinary}
The average parallax of WASP-180 measured by {\it Gaia} DR2 is 3.885\,mas and the angular separation is 4.854\,$''$, which indicates a projected binary separation of $\sim$\,1200\,AU. This would imply an orbit of $\sim$\,30\,000 yrs, which is compatible with the fact that no significant change in separation or position angle is seen in measurements taken over a period of 120 years, as listed in the WDS.

\subsection{Correcting for dilution}
Our photometry of WASP-180 was all extracted from an aperture including both A and B components. Thus we need to correct the lightcurves for dilution.  We deduced correction factors in the different bands of SDSS z and Johnson V, the latter of which was used as an approximation for the WASP data. These are estimated from deducing the effective temperatures of the two stars from available photometry, as follows.

We fitted \teff, \logg, and [Fe/H] by comparing resolved catalogue photometry to the synthetic photometry of \citet{2014MNRAS.444..392C,2018MNRAS.475.5023C} which uses the {\sc marcs} stellar models of \citet{2008A&A...486..951G}. The stars were assumed to have identical [Fe/H]. Interstellar reddening was found to be poorly constrained by the photometry, and was instead fixed at E(B--V)\,=\,0.01, derived from the 3D dust map of \citet{2014ApJ...783..114G,2015ApJ...810...25G}, adopting the closest reliable reddening measurements in the map, at approximately 400\,pc. The choice of distance does not significantly affect the results, with the full line-of-sight reddening out to 8\,kpc being E(B--V)\,=\,0.02\,$\pm$\,0.02. Resolved photometry was found in PANSTARRS-1 \citep[{\it grizy}][]{2016arXiv161205560C}, CMC15 \citep[{\it r'}][]{2014yCat.1327....0N}, DENIS \citep[{\it IJK}][]{1997Msngr..87...27E}, and 2MASS \citep[{\it JHK}][]{2003yCat.2246....0C}. The PANSTARRS-1 catalogue does not include uncertainties for the measurements, and so a conservative uncertainty of 0.1\,mag was assigned to all measurements in that catalogue.

Stellar parameters were derived by least-squares minimisation to find the minimum $\chi^{2}$, and uncertainties were determined by perturbing each parameter separately until a $\delta\chi^{2}$ of 1 was reached. We found \logg\ to be poorly constrained by the photometry, with the entire range of the synthetic photometry grids (3.0\,$\leq$\,\logg\,$\leq$\,5.0) failing to give $\delta\chi^{2}$\,>\,1. Temperatures of 6540$^{+80}_{-30}$\,K and 5430$^{+30}_{-25}$\,K were obtained for the A and B components respectively, as well as a joint [Fe/H] value of 0.0$^{+0.1}_{-0.5}$. The fitting was also repeated four further times, excluding each of the four photometric catalogues (PANSTARRS-1, CMC15, DENIS, 2MASS) in turn. The mean and standard deviation of the parameters from these four additional fits are T$_{\rm A}$\,=\,6521\,$\pm$\,56\,K, T$_{\rm B}$\,=\,5425\,$\pm$\,17\,K, and [Fe/H]\,=\,--0.01\,$\pm$\,0.01: in good agreement with the full fit, indicating that none of the four photometric surveys is significantly biased. The values of T$_{\rm A}$ we obtain are consistent with the value of \teff\ from the spectral analysis (Sec.~\ref{sec:specanalysis}).

Using the stellar parameters from the full fit, and a fixed \logg\ of 4.5 (consistent with spectral analysis), flux ratios were estimated from the synthetic photometry for the z' and V bands. The fraction of light contributed by the secondary star was calculated, and thus the light curves corrected for the dilution of the planetary transit. The third light values and stellar flux ratios we obtained are given in Table~\ref{table:dilution}.

\begin{table}
\caption{Third light dilution factors and stellar flux ratios obtained for WASP-180.}
\centering
\begin{tabular}{lcc}
\hline
Passband & Third Light & Flux Ratio \\[0.5ex]
\hline
SDSS z & 0.325\,$\pm$\,0.007 & 0.48\,$\pm$\,0.01 \\
Johnson V & 0.260\,$\pm$\,0.006 & 0.351\,$\pm$\,0.008 \\
\hline
\end{tabular}
\label{table:dilution}
\end{table}

\subsection{IRFM analysis}
\label{sec:IRFM}
We use the InfraRed Flux Method \citep[IRFM][]{1977MNRAS.180..177B} to derive stellar angular diameters and IR temperatures for \widA\ and \widB. The IRFM makes use of the insensitivity of stellar surface flux to \teff\ at IR wavelengths to determine \teff\ from the ratio of total integrated flux to monochromatic flux, and thus measure the angular diameter of a star. We combine the angular diameters with the {\it Gaia} DR2 parallaxes for the two stars, applying the correction to {\it Gaia} DR2 parallaxes suggested by \citet{2018ApJ...862...61S}, to estimate their radii. We calculate $R_{\star\rm, A}$\,=\,1.17\,$\pm$\,0.08$R_{\odot}$ and $R_{\star\rm, B}$\,=\,1.07\,$\pm$\,0.06$R_{\odot}$.

\subsection{Rotational modulation search}
\begin{figure}
\hspace*{2mm}\includegraphics[width=0.49\textwidth]{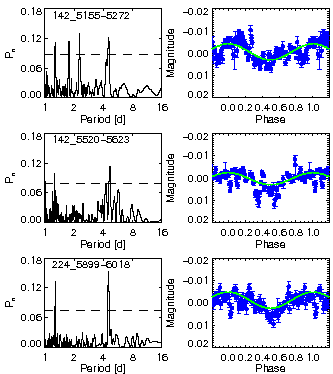}\\ [-2mm]
\caption{Results of the rotational modulation search of the WASP photometry of WASP-180. The three rows show the periodogram (left) and phase-folded light curve (right) for each chunk of data, displayed in the same order as they are listed in Table~\ref{table:rotmodn}. The horizontal dashed line in each of the periodograms corresponds to a confidence level of 99\%.}
\label{fig:rotnmod}
\end{figure}

We perform a search of the WASP photometry following the method of \citet{2011PASP..123..547M}, looking for rotational modulation or pulsation signals with frequencies of 0--1 cycles day$^{-1}$. The data were split into three parts according to the observing season and camera used. We find a signal with an average amplitude of $\sim$\,0.004\,mag and an average period of 4.57\,$\pm$\,0.05\,days. The strongest peak in the first set of data lies at half the modulation period $P_{\rm mod}$. The last set of data contained the clearest signal, and so was given double weight when computing the average. We display the periodograms for each set of data in Fig.~\ref{fig:rotnmod}, and give the individual best-fit amplitudes and periods in Table~\ref{table:rotmodn}.

\begin{table} 
\caption{The results of the rotational modulation search of the WASP photometry of WASP-180. The strongest peak in the periodogram for the first set of data lies at $P_{\rm mod}$/2 (see Fig.~\ref{fig:rotnmod}). The additional peaks around 1--2\,days are ascribed to a combination of harmonics of the rotation period and 1-day aliases.}
\label{table:rotmodn}
\centering
\begin{tabular}{lcccc}
\hline
Dates (HJD-- & No. pts & Period & Amplitude & False Alarm \\
2450000) & & (days) & (mag) & Probability \\
\hline
5155--5272 & 3660 & 2.28 & 0.004 & 0.064 \\
5520--5623 & 3744 & 4.68 & 0.003 & 0.099 \\
5899--6018 & 3171 & 4.53 & 0.004 & <\,0.001 \\
\hline
\end{tabular} 
\end{table}

Using the measured \vsini\ from spectral analysis (18.3\,$\pm$\,1.1\,km\,s$^{-1}$) and the adopted stellar radius from the combined analysis (1.19\,$\pm$0.06\,R$_{\odot}$), we obtain an upper limit on the rotation period of \widA, finding \prot\,<\,3.3\,days. This compares with the modulation period of $\sim$\,4.6\,days, implying that the signal does not originate from rotational modulation in \widA.

The co-moving companion star \widB\ contributes $\sim$\,30\% of the total flux, and so the true amplitude of the signal if originating from the secondary would be $\sim$\,1\%, which is consistent with spot modulation on a fast-rotating later-type star. {\it Gaia} DR2 does not find any other close neighbours which may contribute to the total flux. Thus we believe the signal to belong to the visual companion star, which has a temperature of 5430$^{+30}_{-25}$\,K. A rotation period of 4.6\,days is fairly rapid for a star of \teff\,=\,5430$^{+30}_{-25}$\,K, which may imply a young age for the system, consistent with our {\sc bagemass} analysis in Section~\ref{sec:bagemass}.

\section{Combined MCMC analysis}
\label{sec:combanalysis}
We use a Markov Chain Monte Carlo (MCMC) approach to fit the combined photometric and radial velocity data, as well as investigate the RM effect.  We follow methods very similar to \citet{2018MNRAS.480.5307T,2019AJ....157..141T}, whereby we conduct both an RM analysis and a tomographic analysis and adopt the better-constrained solution. The RM analysis involves detecting the line-profile distortions as an apparent overall shift in radial velocity measurements \citep[e.g.][]{2017haex.bookE...2T}, whereas the tomographic analysis requires one to directly map the motion of the distortion caused by the occulting body across the line profiles as a function of phase \citep[e.g.][]{2017MNRAS.464..810B,2017MNRAS.471.2743T}.

\begin{figure}
\hspace*{2mm}\includegraphics[width=0.49\textwidth]{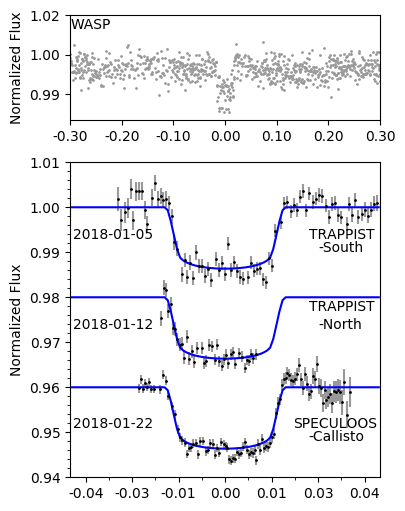}\\ [-2mm]
\caption{The WASP discovery photometry (top) and follow-up transit lightcurves (bottom) with the best-fitting model shown in blue. (see Section~\ref{sec:combanalysis}). The data for the three follow-up lightcurves, prior to the dilution correction, are available online as supporting material.}
\label{fig:phot}
\end{figure}

The code we use is described by \citet{2007MNRAS.380.1230C,2008MNRAS.385.1576P,2010MNRAS.403..151C}. The combined photometric and RV fitting determines the orbital period $P$, the epoch of mid-transit $T_{\rm c}$, the planet-to-star area ratio $(R_{\rm p}/R_{\star})^{2}$, the transit duration $T_{\rm 14}$, the impact parameter $b$, the stellar reflex velocity semi-amplitude $K_{\rm 1}$ and the barycentric system velocity $\gamma$. We use the value of \teff\ obtained in the dilution correction as input, and interpolate four-parameter limb darkening coefficients from the \citet{2000A&A...363.1081C,2004A&A...428.1001C} tables in each step using the current value of \teff. We use the stellar radius obtained in Sec.~\ref{sec:IRFM} (1.17\,$\pm$\,0.08$R_{\odot}$) as a prior to constrain stellar parameters. In the fit we present we have assumed that the orbit is circular, as one would expect a hot Jupiter to circularise on a timescale shorter than its lifetime \citep{2011MNRAS.414.1278P}. However, a further fit was carried out to test this assumption, leading to an upper limit of $e$\,<\,0.27 (95\%\ confidence). We display the photometry and best-fit transit model in Fig.~\ref{fig:phot}.

The RM fit and Doppler tomography give values for \vsini, $\lambda$ and the system $\gamma$-velocity. We use the calibrations of \citet{2011ApJ...742...69H} to fit the RM effect. For Doppler tomography, we assume a Gaussian profile for the perturbation to the stellar-line profiles and fit the intrinsic Full-Width at Half-Maximum (FWHM) of the perturbation, $v_{\rm FWHM}$. Both methods also provide an additional constraint on the impact parameter $b$, although the tomographic method fits this quantity more directly. We estimate the start value for $\gamma$ by fitting a Gaussian profile to the CCFs. We also apply the spectral \vsini\ as a prior in both fitting modes.

\begin{figure}
\hspace*{2mm}\includegraphics[width=0.49\textwidth]{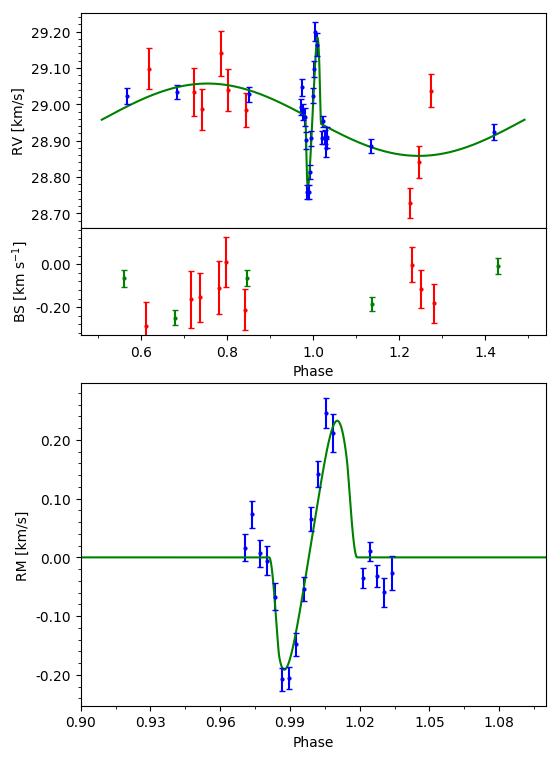}\\ [-2mm]
\caption{Top: All RV measurements of \widA\ used in this work together with the best-fit model shown in green. The red points are CORALIE measurements and the blue points are HARPS measurements. Middle: the bisector (BS) measurements corresponding to the RVs in the top panel. Bottom: The RV measurements taken during transit and best-fit RM model.}
\label{fig:RV-RM}
\end{figure}

We find that the tomographic method was better able to constrain \vsini\ and $\lambda$. In the RM fit, the value of \vsini\ was less constrained, even when using the spectral \vsini\ as a prior. Thus we adopt the solution to the fit including Doppler tomography. We give the solutions for both methods in Table~\ref{table:allResults}. The RV measurements used in this analysis and the best-fit RV and RM models are displayed in Fig.~\ref{fig:RV-RM}.

\begin{figure*}
\hspace*{2mm}\includegraphics[width=0.9\textwidth]{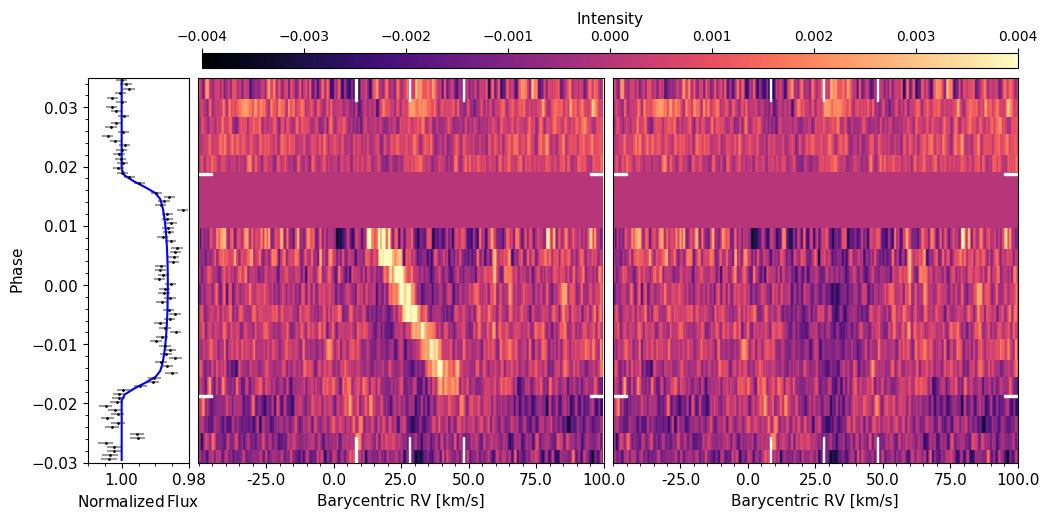}\\ [-2mm]
\caption{The Doppler tomogram for \widb, showing the strong retrograde planet trace (middle) and simultaneous photometric observation alongside (left). The right-hand panel shows the residuals remaining after subtracting the fit to the perturbation due to the planet (see Section~\ref{sec:combanalysis}). The white vertical dashes in the centre and right-hand panels mark the positions of $\gamma$ and $\gamma\,\pm$\,\vsini\, while the white horizontal dashes indicate the times of 1$^{\rm st}$ and 4${\rm th}$ contacts of the planet.}
\label{fig:tomog}
\end{figure*}

Figure~\ref{fig:tomog} shows the tomographic dataset used in this analysis. We have subtracted an average of the out-of-transit CCFs in the dataset from each CCF in order to display the residual bump due to the planet transit. The planet signal is strong and clear, moving in a retrograde direction. Due to excluding three of the CCFs (having low signal-to-noise) we are missing the transit egress. We also show the simultaneous photometric observation in Fig.~\ref{fig:tomog} and a residuals plot produced by subtracting the planet model from the tomographic data.

\begin{table*} 
\caption{All system parameters obtained for WASP-180 in this work. The quantities marked with * were used as priors in the combined MCMC analysis described in Section~\ref{sec:combanalysis}.} 
\label{table:allResults}
\centering
\begin{tabular}{lrrlrr}
\hline
\multicolumn{6}{l}{{\it Stellar system}} \\
\widA\ aliases: & \multicolumn{2}{l}{1SWASP\,J081334.15--015857.9} & \multicolumn{2}{l}{2MASS\,08133416--0158579} & TIC ID:178367144 \\
\widA\ Coordinates: & \multicolumn{5}{l}{RA\,=\,08$^{\rm h}$13$^{\rm m}$34.15$^{\rm s}$\,\,\,\,\,\,Dec\,=\,--01$^{\circ}$58$^{'}$57.9$^{''}$ (J2000)} \\
\multicolumn{6}{l}{Magnitude measurements:} \\
 & \multicolumn{2}{c}{\widA} & \multicolumn{3}{c}{\widB} \\
$B$ (ucac4rpm) & \multicolumn{2}{c}{11.221 $\pm$ 0.3} & \multicolumn{3}{c}{12.732 $\pm$ 0.3} \\
$V$ (ucac4rpm) & \multicolumn{2}{c}{10.682 $\pm$ 0.3} & \multicolumn{3}{c}{12.041 $\pm$ 0.3} \\
$g'$ (Pan-STARRS) & \multicolumn{2}{c}{10.96 $\pm$ 0.3} & \multicolumn{3}{c}{12.336 $\pm$ 0.3} \\
$r'$ (Pan-STARRS) & \multicolumn{2}{c}{10.791 $\pm$ 0.3} & \multicolumn{3}{c}{11.887 $\pm$ 0.3} \\
$i'$ (Pan-STARRS) & \multicolumn{2}{c}{10.786 $\pm$ 0.3} & \multicolumn{3}{c}{11.713 $\pm$ 0.3} \\
$z'$ (Pan-STARRS) & \multicolumn{2}{c}{10.836 $\pm$ 0.3} & \multicolumn{3}{c}{11.637 $\pm$ 0.3} \\
$G$ (Gaia DR2) & \multicolumn{2}{c}{10.9134 $\pm$ 0.0007} & \multicolumn{3}{c}{11.7712 $\pm$ 0.0008} \\
$J$ (2MASS) & \multicolumn{2}{c}{10.11\,$\pm$\,0.05} & \multicolumn{3}{c}{10.68\,$\pm$\,0.03} \\
\hline
\multicolumn{6}{l}{{\it SED analysis}} \\
\teff \medskip & \multicolumn{2}{c}{6540$^{+80}_{-30}$\,K*} & \multicolumn{3}{c}{5430$^{+30}_{-25}$\,K} \\
$\rm [Fe/H]$ \medskip & \multicolumn{2}{c}{0.0$^{+0.1}_{-0.5}$} & \multicolumn{3}{c}{0.0$^{+0.1}_{-0.5}$} \\
\hline
\multicolumn{6}{l}{{\it IRFM, distance and proper motions}} \\
\teff & \multicolumn{2}{c}{6530\,$\pm$\,190\,K}  & \multicolumn{3}{c}{5450\,$\pm$\,130\,K} \\  
$\theta$ & \multicolumn{2}{c}{0.040\,$\pm$\,0.002\,mas}  & \multicolumn{3}{c}{0.038\,$\pm$\,0.004\,mas} \\
\multicolumn{6}{l}{{\it Gaia} DR2 Proper Motions:}\\
RA & \multicolumn{2}{c}{--14.05\,$\pm$\,0.09\,mas\,yr$^{-1}$} & \multicolumn{3}{c}{--12.7\,$\pm$\,0.2\,mas\,yr$^{-1}$} \\
DEC & \multicolumn{2}{c}{--3.17\,$\pm$\,0.06\,mas\,yr$^{-1}$} & \multicolumn{3}{c}{--2.7\,$\pm$\,0.1\,mas\,yr$^{-1}$} \\
{\it Gaia} DR2 Parallax & \multicolumn{2}{c}{3.909\,$\pm$\,0.052\,mas} & \multicolumn{3}{c}{3.862\,$\pm$\,0.073\,mas} \\
$R_{\star}$ & \multicolumn{2}{c}{1.17\,$\pm$\,0.08$R_{\odot}$*} & \multicolumn{3}{c}{1.07\,$\pm$\,0.06$R_{\odot}$} \\
\hline
\multicolumn{6}{l}{\it{Stellar parameters of \widA\ from spectral analysis:}} \\[0.5ex]
Parameter & \multicolumn{2}{c}{Value} & Parameter & \multicolumn{2}{c}{Value} \\
(Unit) & & & (Unit) & & \\
\teff\ (K) & \multicolumn{2}{c}{6500\,$\pm$\,150} & \vsini\ (km\,s$^{-1}$) & \multicolumn{2}{c}{18.3\,$\pm$\,1.1*} \\
\logg\ & \multicolumn{2}{c}{4.5\,$\pm$\,0.2} & $\rm [Fe/H]$ & \multicolumn{2}{c}{0.09 $\pm$ 0.19} \\
$v_{\rm mac}$ & \multicolumn{2}{c}{5.8} & Spectral type & \multicolumn{2}{c}{F7 V} \\
$v_{\rm mic}$ \medskip & \multicolumn{2}{c}{1.5} & \multicolumn{3}{c}{-} \\
\multicolumn{6}{l}{\it{Parameters from combined analyses:}} \\[0.5ex]
Parameter & DT Value & RM Value: & Parameter & DT Value & RM Value: \\
(Unit) & (adopted): & & (Unit) & (adopted): & \\
$P$ (d) & 3.409264 $\pm$ 0.000001 & 3.409265 $\pm$ 0.000001 & \teff\ (K) & 6600 $\pm$ 200 & 6600 $\pm$ 100 \\
$T_{\rm c}$ (BJD$_{\rm TDB}$) & 2457763.3150 $\pm$ 0.0001 & 2457763.3148 $\pm$ 0.0003 &	{[Fe/H]} & 0.1 $\pm$ 0.2 & 0.1 $\pm$ 0.2 \\
$T_{\rm 14}$ (d) & 0.1299 $\pm$ 0.0004 & 0.1285 $\pm$ 0.0009 & $M_{\rm P}$ ($M_{\rm Jup}$) & 0.9 $\pm$ 0.1 & 0.9 $\pm$ 0.2 \\
$T_{\rm 12}=T_{\rm 34}$ (d) & 0.0141 $\pm$ 0.0002 & 0.0145 $\pm$ 0.0008 & $R_{\rm P}$ ($R_{\rm Jup}$) & 1.24 $\pm$ 0.04 & 1.28 $\pm$ 0.09 \\
$R_{\rm P}^{2}$/R$_{*}^{2}$ & 0.0123 $\pm$ 0.0002 & 0.0125 $\pm$ 0.0002 & $\log g_{\rm P}$ (cgs) & 3.12 $\pm$ 0.05 & 3.10 $\pm$ 0.06 \\
$b$ & 0.29 $\pm$ 0.02 & 0.34 $\pm$ 0.06  & $\rho_{\rm P}$ ($\rho_{\rm J}$) & 0.46 $\pm$ 0.05 & 0.43 $\pm$ 0.07 \\
$i$ ($^\circ$) & 88.1 $\pm$ 0.1 & 87.8 $\pm$ 0.4  & $K_{\rm 1}$ (km s$^{-1}$) & 0.10 $\pm$ 0.01 & 0.10 $\pm$ 0.01 \\
$a$ (AU)  & 0.048 $\pm$ 0.001 & 0.049 $\pm$ 0.004 & $T_{\rm P, A=0}$ (K) & 1540 $\pm$ 40 & 1560 $\pm$ 40 \\
$M_{\rm *}$ ($M_{\rm \odot}$) & 1.3 $\pm$ 0.1 & 1.3 $\pm$ 0.3 & \vsini\ (km\,s$^{-1}$) & 19.9 $\pm$ 0.6 & 20.8 $\pm$ 1.5 \\
$R_{\rm *}$ ($R_{\rm \odot}$) & 1.19 $\pm$ 0.06 & 1.17 $\pm$ 0.08  & $\lambda$ ($^\circ$) & --157 $\pm$ 2 & --162 $\pm$ 5 \\
\logg\ (cgs) & 4.42 $\pm$ 0.01 & 4.42 $\pm$ 0.04 & $\gamma$ (km\,s$^{-1}$) & 28.9 $\pm$ 0.1 & 29.0 $\pm$ 0.1 \\
$\rho_{\rm *}$ ($\rho_{\rm \odot}$) & 0.83 $\pm$ 0.01 & 0.82 $\pm$ 0.06 & $v_{\rm FWHM}$ (km\,s$^{-1}$) & 7.9 $\pm$ 0.2 & -- \medskip \\
\hline
\end{tabular} 
\end{table*}

\section{System age determination}
\label{sec:bagemass}
\begin{figure}
\centering
\hspace*{2mm}\includegraphics[width=0.49\textwidth]{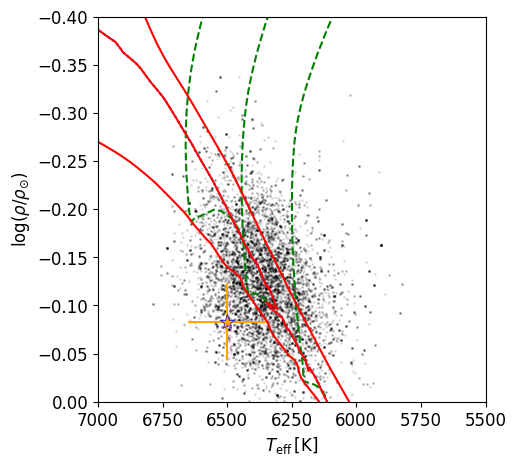}\\ [-2mm]
\caption{The best fitting evolutionary tracks and isochrones of \widA\ obtained using {\sc bagemass}. Black points: individual steps in the MCMC. Dotted blue line: Zero-Age Main Sequence (ZAMS) at best-fit [Fe/H]. Green dashed lines: evolutionary track for the best-fit [Fe/H] and mass, plus $1\sigma$ bounds. The lower-limit evolutionary track lies on top of the ZAMS, making it difficult to see. The Red lines: isochrone for the best-fit [Fe/H] and age, plus $1\sigma$ bounds. Orange star: measured values of \teff\ and $\rho_{*}$ for \widA\ obtained in the spectral and photometric analyses respectively.}
\label{fig:trho_plot}
\end{figure}

We used the open source software {\sc bagemass}\footnote{\url{http://sourceforge.net/projects/bagemass}} to determine the age of the system following a Bayesian approach as described by \citet{2015A&A...575A..36M}. {\sc bagemass} takes constraints on the stellar temperature, density and metallicity to fit the age, mass and initial metallicity of a star using the {\sc garstec} stellar evolution code \citep{2008Ap&SS.316...99W}. We set \teff\,=\,6500\,$\pm$\,150\,K and [Fe/H]\,=\,0.09\,$\pm$\,0.19 (from spectral analysis) and $\rho_{*}/\rho_{\odot}$\,=\,0.83\,$\pm$\,0.01 (from photometry), and use different combinations of mixing lengths and He abundances. We find that the best-fitting parameter set was obtained when using a solar He abundance and mixing length, and thus adopt that solution. We give this solution in Table~\ref{table:agemass} while displaying the evolutionary tracks, isochrones and the distribution of explored values for this fit in Fig.~\ref{fig:trho_plot}. We find \widA\ to be consistent with being on the main sequence, with an age of 1.2\,$\pm$\,1.0\,Gyr. From the best-fit evolutionary tracks we determine the expected main sequence lifetime of the star, taken to be the point at which \widA\ has depleted all hydrogen in the core, is 4.17$^{+0.09}_{-0.71}$\,Gyr.

\begin{figure}
\centering
\hspace*{2mm}\includegraphics[width=0.49\textwidth]{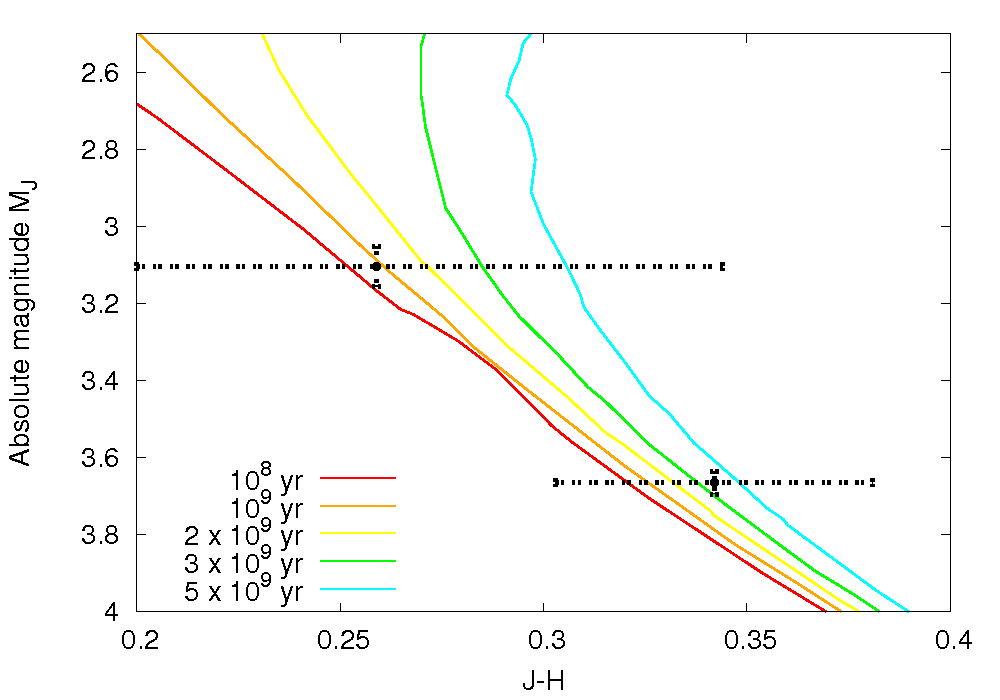}
\caption{A colour-magnitude diagram showing the positions of \widA\ and its comoving companion star with respect to isochrones from \citet{2017ApJ...835...77M} for the ages 0.1, 1, 2, 3 and 5 Gyr (Z = 0.024\,$\sim$\,[Fe/H]\,=\,0.09, Y=0.27).}
\label{fig:isochrones}
\end{figure}

We also extract stellar isochrones from \citet{2017ApJ...835...77M} for stellar ages in the range 10$^{8}$--5$\times\,10^{9}$\,yr, using the metallicity from spectral analysis ([Fe/H]\,$\sim$\,0.09) to estimate appropriate mass fractions, obtaining Z\,=\,0.024 and Y\,=\,0.27. These are displayed on a colour--magnitude diagram in Fig.~\ref{fig:isochrones} along with the positions of \widA\ and \widB. The position of \widA\ implies a system age of $\sim$\,1\,Gyr while the position of \widB\ implies an age close to $\sim$\,3\,Gyr. The positions of \widA\ and \widB\ in Fig.~\ref{fig:isochrones} imply approximate stellar masses of $\sim$\,1.3\,$M_{\star}$ and $\sim$\,1.0\,$M_{\star}$ respectively, leading to a mass ratio of $M_{\star,B}$/$M_{\star,A}$\,$\approx$\,0.77.

\begin{table} 
\caption{Results for the masses of \widA\ and \widB , and the age of the system.} 
\label{table:agemass}
\centering
\begin{tabular}{lc}
\hline
\multicolumn{2}{l}{\it{Parameters from} {\sc bagemass}:} \\[0.5ex]
Parameter & Value \\
(Unit) & \\
Age & 1.22 $\pm$ 0.99 \\
$M_{\star,A}$ ($M_{\rm \odot}$) & 1.18 $\pm$ 0.08 \\
$\rm [Fe/H]_{\rm init}$ & --0.06 $\pm$ 0.16 \\
\hline
\multicolumn{2}{l}{\it{Parameters from stellar isochrones}:} \\
$M_{\star,A}$ ($M_{\rm \odot}$) & 1.3 \\
$M_{\star,B}$ ($M_{\rm \odot}$) & 1.0 \\
\hline
\end{tabular} 
\end{table}

\section{Conclusions and discussion}
\widb\ is a 0.9\,$\pm$\,0.1\,$M_{\rm Jup}$, 1.24\,$\pm$\,0.04\,$M_{\rm Jup}$ hot Jupiter orbiting an F7 V star with 
\teff\,=\,6500\,K and \vsini\,=\,19.9\,km\,s$^{-1}$. The planet's large radius is in line with the expectation for a Jovian-mass planet in a close orbit around a fairly hot star to be inflated due to the high level of irradiation  \citep[e.g.][]{2012A&A...540A..99E,2018A&A...616A..76S}.

The orbit is misaligned and retrograde, with a projected obliquity of $\lambda$\,=\,--157\,$\pm$\,2\,$^{\circ}$. This is also in line with known trends amongst hot Jupiters orbiting hot stars, since the majority of such planets are found to be in misaligned orbits \citep[e.g.][]{2010ApJ...718L.145W,2012ApJ...757...18A,2017AJ....153..205D,2017haex.bookE...2T}.

WASP-180 is a known binary system. We can ask whether the secondary, \widB, is responsible for the retrograde, misaligned orbit seen in \widb, through having induced Lidov-Kozai oscillations leading to high-eccentricity migration of the planet. While this effect has long been thought able to produce such orbits, the pathways leading from high-eccentricity migration to the observed distribution of system obliquities are still a topic of avid research \citep[e.g.][]{2016MNRAS.456.3671A,2017MNRAS.465.3927S}. \citet{2016MNRAS.456.3671A} places an upper limit on the final period of a hot Jupiter which has migrated due to Lidov-Kozai oscillations of \porb\,<\,4\,days, while \citet{2015ApJ...799...27P} finds that the stellar separations of binaries with hot Jupiters are preferentially in the range 400--1500\,AU, and so with \porb\,=\,3.4\,days and an estimated stellar separation of 1200\,AU it is feasible for \widb\ to have formed in this way. \citet{2016MNRAS.456.3671A} also finds that the expected timescale required for the migration of a hot Jupiter of 1\,M$_{\rm Jup}$ via Lidov-Kozai oscillations is in the range $\sim$0.5--5\,Gyr, with lower mass planets taking longer to migrate. The system age of 1.2\,$\pm$\,1.0\,Gyr is consistent with being within this range. It is possible that some eccentricity could remain, however, and our measured upper limit of $e$\,<\,0.27 at 95\%\ confidence implies a possibly eccentric, but likely near circular orbit.
%This is basically waffle since you give no timescale.  What is the timescale for  circularisation?  Is there actually a discrepancy between it having circularised and your age estimate?  

\widb\ has \teff\,=\,6500\,K and \vsini\,=\,19.9\,$\pm$\,0.6\,km\,s$^{-1}$. Another example of a hot Jupiter in a binary system with an early-type host star is KELT-19Ab, with \teff\,=\,7500\,K and \vsini\,=\,84\,$\pm$\,2\,km\,s$^{-1}$ \citep{2018AJ....155...35S}. KELT-19Ab has a measured obliquity of $\lambda$\,=\,--179$^{\circ}$ and so is also on a retrograde orbit. KELT-19Ab is also similar to \widb\ in that the primary and secondary stars in the system are of similar brightness. Such systems are rare, likely due to selection bias. Others include K2-29b \citep{2016ApJ...824...55S} and HAT-P-20b \citep{2011ApJ...742..116B}.

Also, \citet{2015ApJ...800..138N,2015ApJ...814..148P,2016ApJ...827....8N} studied known exoplanet systems with FGK host stars, searching for previously unseen stellar companions and attempting to find a correlation between the presence of a distant stellar companion and the measured obliquity and eccentricity of a hot Jupiter's orbit. They find no evidence of such a trend and conclude that, although a significant fraction of hot Jupiters reside in wide binary systems, fewer than 20\% of hot Jupiters could have ended up in their current orbits as a result of Lidov-Kozai oscillations. Although both KELT-19Ab and \widb\ are in misaligned, retrograde orbits, this is not necessarily related to the fact that they are in binary systems, since the tendency for hot Jupiters orbiting hot stars to be misaligned is well established \citep{2010ApJ...718L.145W,2012ApJ...757...18A}. % Why fewer than 20 %?  because of a lack of companions?

The strength of the planet signal in tomography for \widb\ makes it a potential candidate for looking for differential rotation following the RM reloaded technique of \citet{2016A&A...588A.127C}, through which the effects of differential rotation and the perturbation due to the planet can be disentangled.  To bring out the effect of differential rotation on the spectroscopic transit more clearly, the higher spectral resolution and greater light collecting power of ESPRESSO on the VLT would be of use.

\section*{Acknowledgements}
WASP-South is hosted by the South African Astronomical Observatory and we are grateful for their ongoing support and assistance. Funding for WASP comes from consortium universities and from the UK's Science and Technology Facilities Council. The research leading to these results has received funding from the European Research Council (ERC) under the FP/2007-2013 ERC grant agreement no. 336480, and under the H2020 ERC grant agreement no. 679030; and from an Actions de Recherche Concert\'ee (ARC) grant, financed by the Wallonia-Brussels Federation. The Euler Swiss telescope is supported by the Swiss National Science Foundation (SNF). TRAPPIST-South is funded by the Belgian Fund for Scientific Research (Fond National de la Recherche Scientifique, FNRS) under the grant FRFC 2.5.594.09.F, with the participation of the SNF. M. Gillon and E. Jehin are F.R.S.-FNRS Senior Research Associates. We acknowledge use of the ESO 3.6-m/HARPS spectrograph under program 0100.C-0847(A), PI C. Hellier. This work has made use of data from the European Space Agency (ESA) {\it Gaia} mission (\url{https://www.cosmos.esa.int/gaia}), processed by the {\it Gaia} Data Processing and Analysis Consortium (DPAC, \url{https://www.cosmos.esa.int/web/gaia/dpac/consortium}). Funding for the DPAC has been provided by national institutions, in particular the institutions participating in the {\it Gaia} Multilateral Agreement. This research has also made use of: the NASA Exoplanet Archive, which is operated by the California Institute of Technology, under contract with the National Aeronautics and Space Administration under the Exoplanet Exploration Program; the CMC15 Data Access Service at CAB (CSIC-INTA).
%%%%%%%%%%%%%%%%%%%% REFERENCES %%%%%%%%%%%%%%%%%%

\bibliographystyle{mnras}
\bibliography{litbiblio}

\bsp
\label{lastpage}
\end{document}